\documentclass[structabstract]{aa}

\usepackage{epsfig}
\usepackage{graphicx}
\usepackage[latin1]{inputenc}
\usepackage{txfonts}
\usepackage{rotating}
\usepackage{natbib}

\begin{document}

   \title{Probing isolated compact remnants with microlensing}

   \author{N. Sartore
          \inst{1},
          A. Treves
          \inst{1,2} }

   \institute{Dipartimento di Fisica e Matematica, Università dell'Insubria,
              via Valleggio 11, 22100, Como, Italy \\
              \email{nicola.sartore@gmail.com}
              \and 
              Affiliated to INAF and INFN
             }

   \date{Received ...; accepted ...}

 \abstract
    {}
    {We consider isolated compact remnants (ICoRs), i.e. neutrons stars and black holes that do not reside in binary systems 
    and therefore cannot be detected as X-ray binaries. 
    ICoRs may represent $\sim\,5$ percent of the stellar mass budget of the Galaxy, but they are very hard to detect.
    Here we explore the possibility of using microlensing to identify ICoRs.}
    {In a previous paper we described a simulation of neutron star evolution in phase space in the Galaxy, taking into account the
    distribution of the progenitors and the kick at formation.
    Here we first reconsider the evolution and distribution of neutron stars and black holes adding a bulge component.
    From the new distributions we calculate the microlensing optical depth, event rate and distribution
    of event time scales, comparing and contrasting the case of ICoRs and "normal stars".}
    {We find that the contribution of remnants to optical depth is slightly lower than without kinematics, owing to the evaporation from the Galaxy. 
    On the other hand, the relative contribution to the rate of events is a factor $\sim\,5$ higher. 
    In all, $\sim\,6-7$ percent of the events are likely related to ICoRs. 
    In particular, $\sim\,30-40$ percent of the events with duration $>\,100$ days are possibly related to black holes.}
    {It seems therefore that microlensing observations are a suitable tool to probe the population of Galactic ICoRs.}

   \keywords{stars: kinematics - stars: neutron - stars: statistics}

\maketitle
\section{Introduction}\label{intro}
By isolated compact remnants we mean neutron stars and black holes (NSs and BHs hereafter) that are generated from the core-collapse of isolated massive stars.
Hence, these objects cannot be recycled by accretion of matter stripped from a companion star.
For NSs we consider in particular those objects that exhausted their possible reserves of energy, like rotation and internal heat, and are not detectable as active sources, as in the case of radio pulsars or X-ray pulsars, like the Magnificent Seven (e.g. \citealt{Ha07}, \citealt{Tu09} and references therein).

Extensive searches for ICoRs as sources powered by accretion from the interstellar medium (e.g. \citealt{PP98, Tr00, AK02}) were performed,
for example, in the ROSAT catalog (e.g. \citealt{NT99}).
However, no positive confirmation of an accreting ICoR has been reported to date.

\cite{Of10} suggested that the recently discovered long duration radio transients (e.g. \citealt{Bo07}) may be associated with isolated old neutron stars (ONS).
Hence, a possible direct detection of ICoRs may already be available, but is not yet fully demonstrated.
On the other hand \cite{Be02} suggested that ICoRs, in particular black holes, may be responsible for some of the long-duration microlensing events observed.

Microlensing surveys (see \citealt{Mo10} for a recent review) are used to study the distribution of dark as well as luminous
matter along a given line of sight (l.o.s. hereafter), and therefore this may be a suitable method to probe the phase space distribution of ICoRs. 
The first theoretical studies of the subject (\citealt{Gr91}, \citealt{Pa91}) predicted an optical depth $\tau\,\sim\,0.5 - 0.8\,\times\,10^{-6}$ 
for the luminous stellar component of the Milky Way (see Sect. \ref{model} for a definition of $\tau$).
Several observational campaigns have been set up to detect microlensing events in the direction of the Galactic bulge, e.g.
MACHO (\citealt{Al97}), OGLE (\citealt{U94b}) and MOA (\citealt{Su03}).

The first measurements of the optical depth returned values that are significantly higher than those predicted, e.g
$\tau\,=\,3.3\,\times\,10^{-6}$ (\citealt{U94b}) and $\tau\,=\,3.9\,\times\,10^{-6}$ (\citealt{Al97}). 
To explain this discrepancy, \citealt{KP94} suggested that the Galactic bulge could give a substantial contribution to the optical depth and thus they included this contribution in their calculations.
Theoretical estimates with these new models agree well with the more recent measurements of the optical depth,
$\tau\,=\,2.17\,\times\,10^{-6}$ (\citealt{Po05}, the MACHO group) , $\tau\,=\,2.55\,\times\,10^{-6}$ (\citealt{Su06}, 
the OGLE-2 group) and $\tau\,=\,1.62\,\times\,10^{-6}$ (\citealt{Ha06}, the EROS-2 group).

The contribution of compact remnants to microlensing has been estimated by \citet{Go00, WM05, CN08}. 
These authors found that $\sim\,2 - 3$ percent of the events are related to ICoRs. 
However, they did not take into account that ICoRs have a different phase space distribution with respect to normal stars, 
because of the high kick velocities received at birth, (e.g. \citealt{Ho05, Gu05}).
This has two opposite effects. First, the velocities of ICoRs at birth can be higher than the escape velocity from the Galaxy. 
Hence, a non negligible fraction of remnants may have escaped from the Galaxy and thus cannot contribute to microlensing toward the bulge. 
Second, the higher velocities of ICoRs imply an event rate higher than that expected from a similar population of lenses moving at lower speeds (\citealt{Gr91}). 
Thus, in this paper we investigate the possibility to probe isolated compact remnants by means of microlensing observations. 
We base our work on the results of Monte Carlo simulations of NS orbits (\citealt{Of09}, \citealt{Sa10}, hereafter Of09 and Paper I, respectively). 

The paper is organized as follows. 
In Sect. \ref{model} we describe our models for the distribution of bulge and disk stars as well as ICoRs after an introduction of the basic microlensing quantities and expressions. 
We present our results in Sect. \ref{results}. 
In particular we compare the optical depth and event rate due to ICoRs with that of normal stars. 
We also study the distribution of event time scales in both cases. 
Finally we discuss our results in Sect. \ref{discussion}.

\section{Model}\label{model}
We now introduce some basic expressions related to the microlensing phenomenon. 
The probability that a background source is lensed at any given time is called microlensing optical depth and depends on the
distribution of lensing matter along the l.o.s. (e.g. \citealt{KP94}, \citealt{Je02})

\begin{equation}\label{eqn_tau1}
    \tau(D_s)\,=\,\frac{4\,\pi\,G}{c^2}\int_0^{D_s} \rho_l(D_l) \frac{D_l(D_s -
    D_l)}{D_s}\,\textrm{d}D_l\,,
\end{equation}

\noindent where $\rho_l$ is the mass density of the lenses, $D_s$ is the source-observer distance, and $D_l$ is the lens-observer distance. 
If both lenses and sources are distributed along the l.o.s., Eq. \ref{eqn_tau1} becomes

\begin{equation}\label{eqn_tau2}
    \tau\,=\,\frac{4\,\pi\,G}{c^2\,I}\,\int_0^{D_{max}}\,\tau(D_s)\,\rho_s(D_s)\,D_s^2\,\textrm{d}D_s\,,
\end{equation}

\noindent where $\rho_s$ is the mass density of sources and $I\,=\,\int_0^{D_{max}}\,\rho_s(D_s)D_s^2\textrm{d}D_s$ is a normalization factor.
Throughout this paper we assume $D_{max}\,=\,12$ kpc to include the contribution of the whole bulge (see Sect. \ref{bd_model}). 
The optical depth can also be defined as (\citealt{Gr91})

\begin{equation}\label{eqn_tau3}
    \tau\,=\frac{\pi\,\left\langle t_E \right\rangle\,\Gamma}{2}\,,
\end{equation}

\noindent where $\Gamma$ is the rate of lenses entering the tube and $\left\langle t_E \right\rangle$ is the average time scale of the observed events. 
The duration of a single event depends on the mass of the lens and the geometry of the system

\begin{equation}\label{eqn_re}
    t_E\,=\,\frac{R_E}{v_\bot}\,=\,\frac{2}{v_\bot}\,\Big[ \frac{G\,M}{c^2}\,\frac{D_l\,(D_s - D_l)}{D_s}\,\Big]^{1/2}\,,
\end{equation}

\noindent where $v_\bot$ is the relative velocity between source and lens in a plane perpendicular to the l.o.s. and $R_E$ is the Einstein radius (e.g. \citealt{Gr91})


\begin{eqnarray}\label{eqn_re}
	R_E\,=\,2\,\Big[\frac{G\,M}{c^2}\,\frac{D_l\,(D_s - D_l)}{D_s}\,\Big]^{1/2} \nonumber \\
	\,=\,1.38\,\times\,10^{-8}\,\Big[m\,D_s\,x\,(1 - x)\Big]^{1/2}\,\rm\,kpc\,,
\end{eqnarray}

\noindent where M is the mass of the lens, $x\,=\,D_l/D_s$ and $m\,=\,M\,/\,M_\odot$ is the mass of the lens in solar units.

The differential rate of events is (e.g. \citealt{Je02}, \citealt{CN08})

\begin{equation}\label{eqn_rate1}
    \rm{d}\Gamma\,=\it\,\frac{n_l(D_l)\,\rm{d}^3\it D_l}{\rm{d}t}\,\frac{n_s(D_s)\,D_s^2\,\rm{d} \it D_s}{I}\,f(\bf{v_\bot})\,\rm{d}^2 \it v_\bot\,,
\end{equation}

\noindent where $n_l$ and $n_s$ are the number density of the lenses and the sources along the l.o.s., 
while $f(\bf{v_\bot})$ is the distribution of the source-lens relative velocities.

\subsection{Distribution of bulge and disk stars}\label{bd_model}

We model the Galactic bulge as in \cite{CN08}, i.e. as a triaxial bulge with an exponential density profile and a major axis that forms an angle $\phi\,=\,23.8°$ with the Sun-Galactic center axis (e.g \citealt{St97}). The total mass of the bulge is $\sim\,2\,\times\,10^{10}\,M_\odot$.
For the disk, we adopt a thin + thick disk model, both described by an exponential density profile


\begin{eqnarray}\label{eqn_dend}
    \rho_{D_i}(R,z)\,=\,\frac{M_{D_i}}{4\,\pi\,(L_i^2 - L_h^2)}\,\Big[\exp\Big(\frac{R}{L_i}\big)\,-\,\exp\Big(\frac{R}{L_h}\big)\Big] \nonumber \\
		\times\,\exp\big(-\frac{|z|}{H_i}\big),\:i=1,2\,,
\end{eqnarray}

\noindent where the masses of the thin and thick disks are $M_{D_1}\,\sim\,2.5\,\times\,10^{10}\,M_\odot$ and
$M_{D_2}\,\sim\,0.5\,\times\,10^{10}\,M_\odot$, the scale-lengths are $L_1\,\sim\,2.6$ kpc and $L_2\,\sim\,3.6$ kpc and
scale-heights are $H_1\,\sim\,0.3$ and $H_2\,\sim\,0.9$ kpc (e.g. \citealt{Ro03}).
The mass of the thin disk accounts also for the interstellar medium, $M_{ISM}\,\sim\,0.5\,\times\,10^{10}\,M_\odot$.
According to \cite{Fr98}, the stellar disk is holed at its center, the hole being likely
produced by orbital resonances in the potential of the barred bulge (e.g. \citealt{Co89}). 
For the scale-length of the hole of both the thin and thick disks we assume $L_h\,\sim\,1.3$ kpc \citep{PR04}.

The motion of bulge and disk stars has both bulk and random components. 
The Galactic bulge does not rotate as a whole like a rigid body but out of a certain radius the bulk velocity flattens (\citealt{Ri07}). 
Thus we assume that the rotation velocity $v_{rot}^{(b)}$ of the bulge grows linearly to
$50\,\textrm{km\,s}^{-1}$ out of a radius of 1 kpc from the Galactic center. 
Out of this radius, we assume a flat rotation curve, with $v_{rot}^{(b)}=50\,\rm km\,s^{-1}$. 
For disk stars, we compute the bulk motion self-consistently from the potential generated by disk and bulge stars, see Sect. \ref{sect_icors}. 
We assume for simplicity that the random motions of bulge and disk stars are isotropic, with dispersions $\sigma_v^{(b)}=100$ and $\sigma_v^{(d)}=25\,\rm km\,s^{-1}$ which reasonably agrees with the values inferred from observations (e.g. \citealt{CN08} and references therein).

\subsection{Isolated compact remnants}\label{sect_icors}

In Paper I we studied the dynamics of a population of disk NSs born at a constant rate during the Milky Way
lifetime, assuming a total of $10^9$ NSs, consistently with chemical abundances observed in the Galaxy (e.g. \citealt{Ar89}).
As Of09 has pointed out, the total number of disk-born NSs inferred from the present-day supernova rate and from the star-formation history of the disk is $\lesssim\,4\,\times\,10^8$ (\citealt{KK08} and references therein). 
To explain this discrepancy, Of09 suggested that the remaining NSs have been generated in the bulge.

Because we are dealing with the microlensing rate of compact remnants toward the Galactic bulge, 
it is straightforward to think that the major contribution comes from bulge-born objects. 
Thus we re-run our simulations, taking into account the contribution of the bulge. 
First, we estimate the total number of ICoRs born in the MW, following the approach of \cite{Go00}. 
We adopt the initial mass function (IMF) proposed by \cite{Kr01}, i.e. a triple power law model

\begin{equation}
          \begin{array}{l@{\quad}l@{\quad\quad}l}          
	\frac{\rm{d}N}{\rm{d}m}\,\propto\,m^{-\alpha}\,, & \alpha = 0.3\,, & 0.03 < m < 0.08 \\
 																										& \alpha = 1.3\,, & 0.08 < m < 0.50 \\
																										& \alpha = 2.3\,, & 0.50 < m < 100\,. \\
          \end{array}
\end{equation}

From the IMF we estimate the number and mass fractions of each stellar population - brown dwarfs, main sequence stars, white dwarfs (BDs, MSs and WDs hereafter), NSs and BHs.
We assume that all stars with $m\,>\,1$ have evolved through the remnant phase, i.e. stars with $1\,<\,m\,<\,8$ are now white dwarfs ($m_{WD}\,=\,0.6$), while stars with masses $8\,<\,m\,<\,40$ and $40\,<\,m\,<\,100$ are treated as neutron stars ($m_{NS}\,=\,1.4$) and black holes ($m_{BH}\,=\,10$), respectively. 
Results are reported in Table \ref{table_imf}.
We can now obtain the number of NSs and BHs generated in the Galaxy: $N^{(b)}_{NS}\,\sim\,3.3\,\times\,10^8$, $N^{(b)}_{BH}\,\sim\,3.2\,\times\,10^7$, $N^{(d)}_{NS}\,\sim\,4.1\,\times\,10^8$ and $N^{(d)}_{BH}\sim\,4.0\,\times\,10^7$, where the superscripts \textit{(b)} and \textit{(d)} refer to the bulge and disk populations, respectively.
We note that the assumption that all disk stars with $1\,<\,m\,<\,8$ are now white dwarfs gives only a rough estimate of the number of disk WDs.
On the other hand the assumption is well justified for NSs and BHs, because the typical lifetime of massive stars is much shorter than the age of the Galaxy.

\begin{table}[ht]
	\centering
		\begin{tabular}{c c c c c c}
\hline
 & BD & MS & WD & NS & BH \\
\hline
number fraction & 0.272 & 0.653 & 0.065 & 0.004 & 0.0004 \\
mass fraction & 0.059 & 0.744 & 0.157 & 0.023 & 0.016 \\
\hline
		\end{tabular}
\caption{Number and mass fractions of the different stellar populations.}
\label{table_imf}
\end{table}

We use the PSYCO code (Paper I) to follow the orbits of $2\,\times\,10^5$ bulge-born and $2\,\times\,10^5$ disk-born synthetic ICoRs.
We have revised the Galactic potential assumed in Paper I, considering a superposition of Miyamoto and Nagai disks (\citealt{MN75}) for both the bulge and disk.
The associated density profiles are a close approximation of the density profiles assumed for the bulge and disk of the Galaxy.
As in Paper I, we adjusted the parameters of the dark matter potential to obtain a circular velocity of $\sim\,220\,\rm km\,s^{-1}$ at the solar circle ($R_\odot\,=\,8.5\, \rm kpc$).
The resulting escape velocity at the same radius is $\sim\,450\,\rm km\,s^{-1}$, i.e. $\sim\,17$ percent lower than in Paper I.
Also, we make the simplifying assumption that the gravitational potential of the bulge is axisymmetric. 

The results obtained are then normalized to the corresponding number of objects expected from each population and reported above.
The initial conditions for each object are randomly assigned using a Monte Carlo procedure. 
The integration time for each object is equal to its age, i.e. 10 Gyr for all bulge ICoRs, 
while the age of disk ICoRs is uniformly distributed between 0 and 10 Gyr.
In both cases the radial birth probability is assumed to be proportional to the density of stars, 
and we assume also that the bulge and disk can produce remnants up to 3 and 15 kpc, respectively. 
For the bulge, the vertical birth probability is again assumed proportional to the density of normal stars up to 3 kpc, while for the disk we adopt the same distribution as the one proposed by \cite{Pa90}, which follows the distribution of young stars in the disk of the MW. 
The azimuthal coordinates are uniformly distributed in both cases.

The initial velocity of each synthetic remnant is the vector sum of its birth velocity plus the orbital velocity of the progenitor.
The orbital velocity of the progenitor is calculated from the gravitational potential.
The birth velocities of NSs are distributed according to a Maxwellian distribution with dispersion $\sigma\,=\,265\,\rm{km\,s}^{-1}$ \citep{Ho05}.
The velocity distribution of BHs is poorly constrained, and it was assumed that these objects have a dispersion of $\sim\,40\,\rm{km\,s}^{-1}$ 
(e.g. \citealt{WvP96}).
However, the discovery of BH X-ray binaries with high spatial velocities (e.g. \citealt{Mi01, Mi02}), 
points to a similar birth velocity distribution for NSs and BHs.
Hence, in this work we assume that BHs have the same velocity distribution of NSs.
Finally, we also add a random component to the velocity of the progenitors, see Sect. \ref{bd_model}.
For the bulge, this dispersive component is dominant with respect to the orbital velocity around the Galactic center.

\section{Results}\label{results}

\subsection{Neutron stars}\label{results_ns}
\begin{figure*}
\centering
  \includegraphics[width=0.48\textwidth]{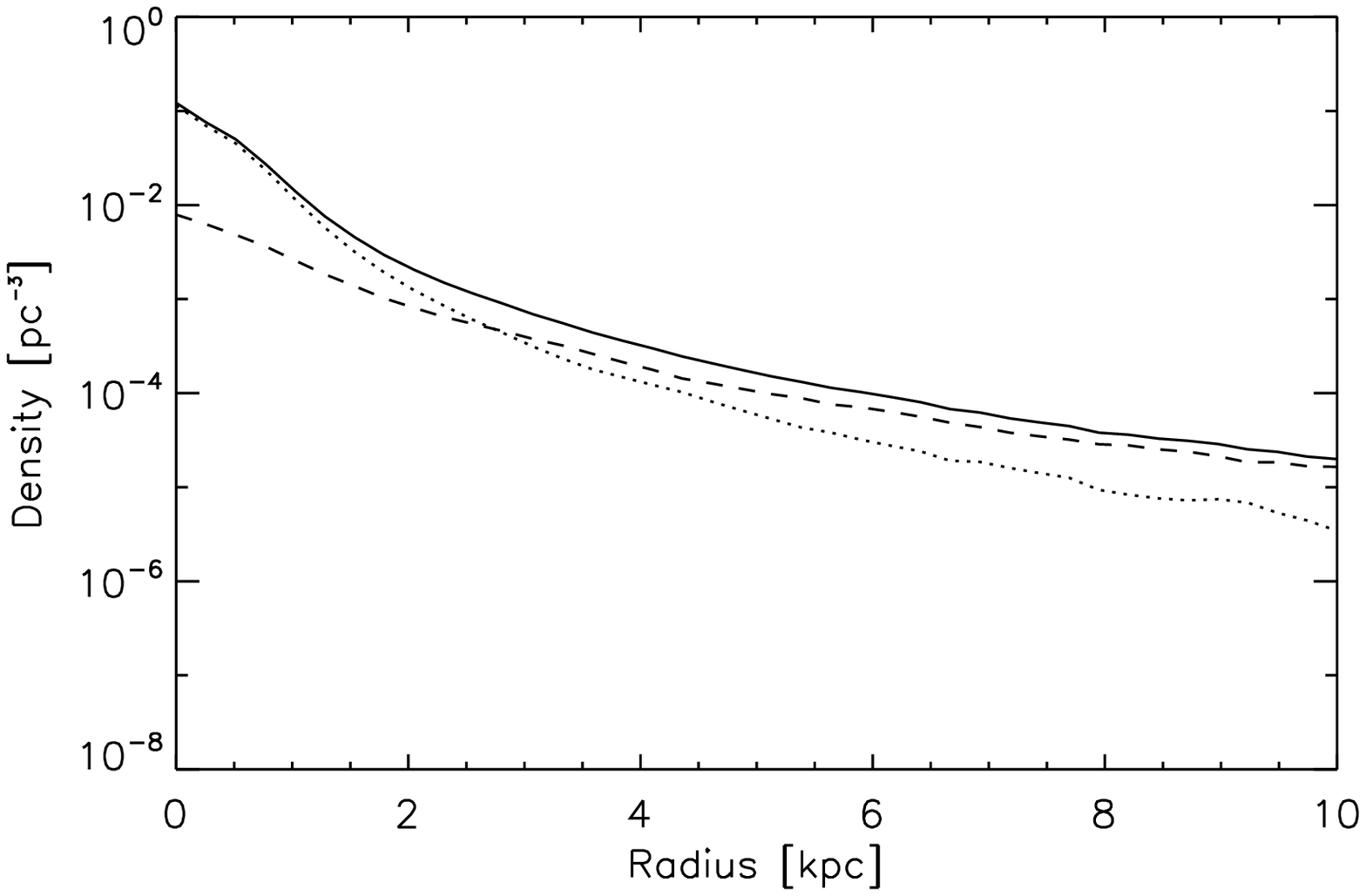}
  \includegraphics[width=0.48\textwidth]{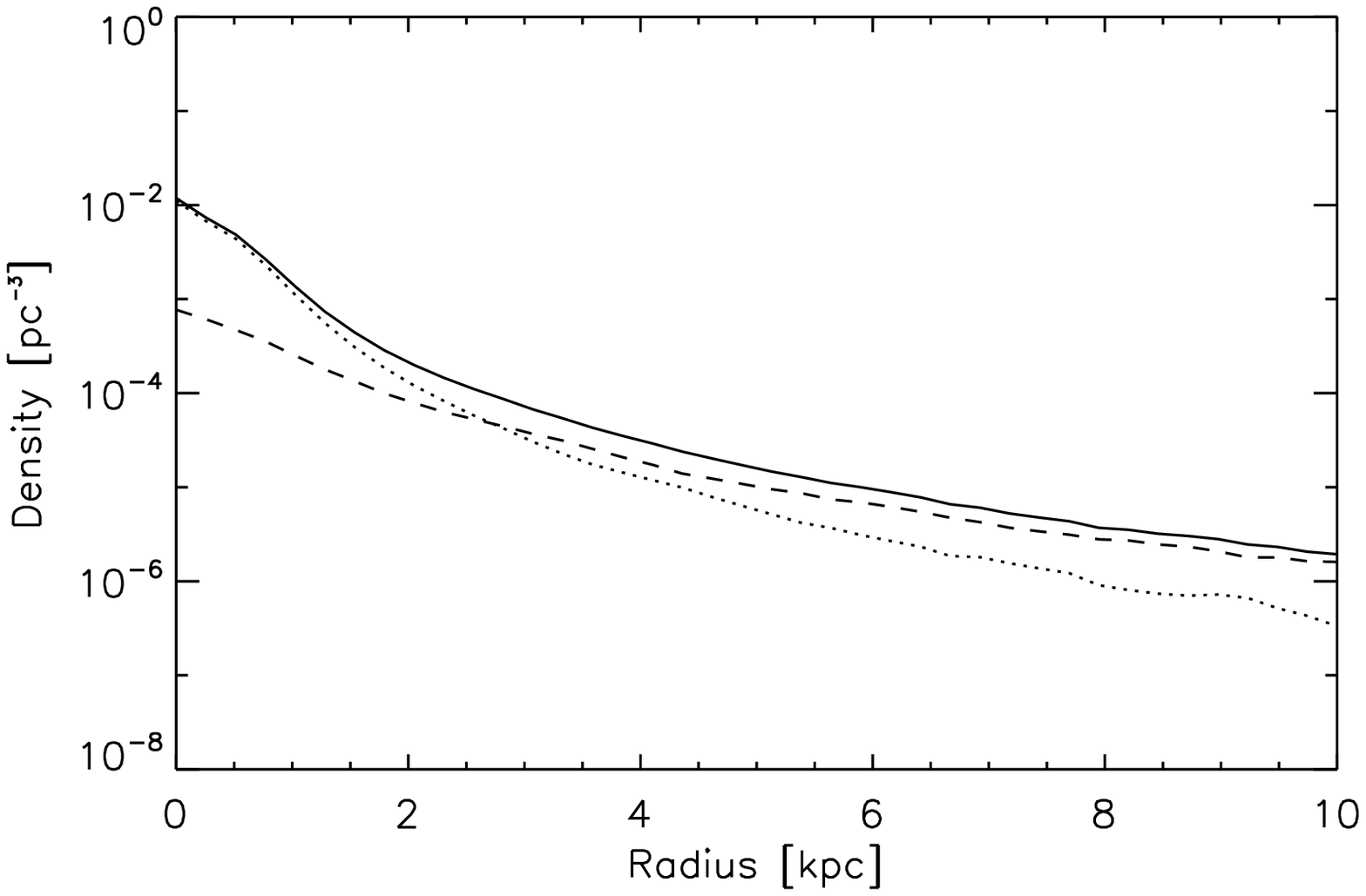}  
  \caption{Density profiles of NSs (left) and BHs (right) in the Galactic plane. 
  	Solid, dotted, and dashed lines represent the total, bulge, and disk contributions.}
  \label{ons_profile}
\end{figure*}

First, we compare the statistical properties of NSs in the new model with those reported in Paper I and Of09.
We find that for bulge NSs, the evaporation is highly inefficient, that is, almost all NSs are in bound orbits, $f_{bound}\,\sim\,0.98$, because of the high gravitational force toward the inner part of the Galaxy.
The fraction of disk-born NSs in bound orbits is slightly lower than that found in Paper I for the same velocity distribution,
$f_{bound}\,\sim\,0.63$ because, as we already pointed out, the escape velocity in the new model is also lower.
In all, $\sim\,20$ percent of NSs evaporated from the Galaxy.
The density of bulge-born NSs becomes dominant in the inner part of the Galaxy, $R\,\lesssim\,3\,\rm kpc$.
In particular the density of ONS in the bulge can be as high as $\sim0.1\,\rm pc^{-3}$, see Fig. \ref{ons_profile} (left panel).

Our results are different from those obtained by Of09, for which $f_{bound}$ varies from 0.41 to 0.52 for bulge NSs and from 0.13 to 0.16 for disk NSs, respectively, depending on the initial velocity distribution.
The main reason is the difference in the gravitational potential adopted for the MW.
In our model the total mass of the Galaxy is $\sim\,10^{12}\,M_\odot$ (e.g. \citealt{Xu08}), i.e. a factor $\sim\,7$ larger than that adopted by Of09.
This obviously implies that in our model a larger fraction of ONS is bound to the Galaxy.
Furthermore, we adopt different initial conditions, e.g. the distribution of progenitors and birth velocities of NSs and 
a different (lower) normalization.

The local density of NSs is $n_0\,\sim\,3.3\,\times\,10^{-5}\,\rm pc^{-3}$, of which the fraction of bulge-born ONS is $\sim\,0.23$.
The density is therefore a factor $\sim\,3$ lower than what we found in Paper I for the same velocity distribution and 
a factor $\sim\,10$ lower than that reported by Of09. 
We note however that in Paper I we obtained a value of the local density similar to that of Of09 when using the same velocity distribution.
Also, the fraction of bulge-born objects is similar to that of Of09.
This means that local density is affected mainly by the normalization, i.e. the total number of NSs in the Galaxy.

We find that the density in the Galactic center is $n_{GC}\,\sim\,0.12\,\rm pc^{-3}$, of which $\sim\,93$ percent have been born in the bulge itself.
These results are similar to those of Of09, who obtained $n_{GC}\,\sim\,0.2\,-\,0.3\,\rm pc^{-3}$, with $95\,-\,97$ percent of NSs being bulge-born.
The larger contribution of bulge NSs to the density toward the Galactic center can be appreciated for example in Fig. \ref{pden}, 
where the projected density within 12 kpc from the Sun is $\sim\,7.3\,\times\,10^4\,\rm deg^{-2}$ against
$\sim\,6.7\,\times\,10^3\,\rm deg^{-2}$ of disk-born NSs.

\subsection{Black holes}\label{results_bh}
\begin{figure*}[ht]
  \centering
  \includegraphics[width=0.45\textwidth]{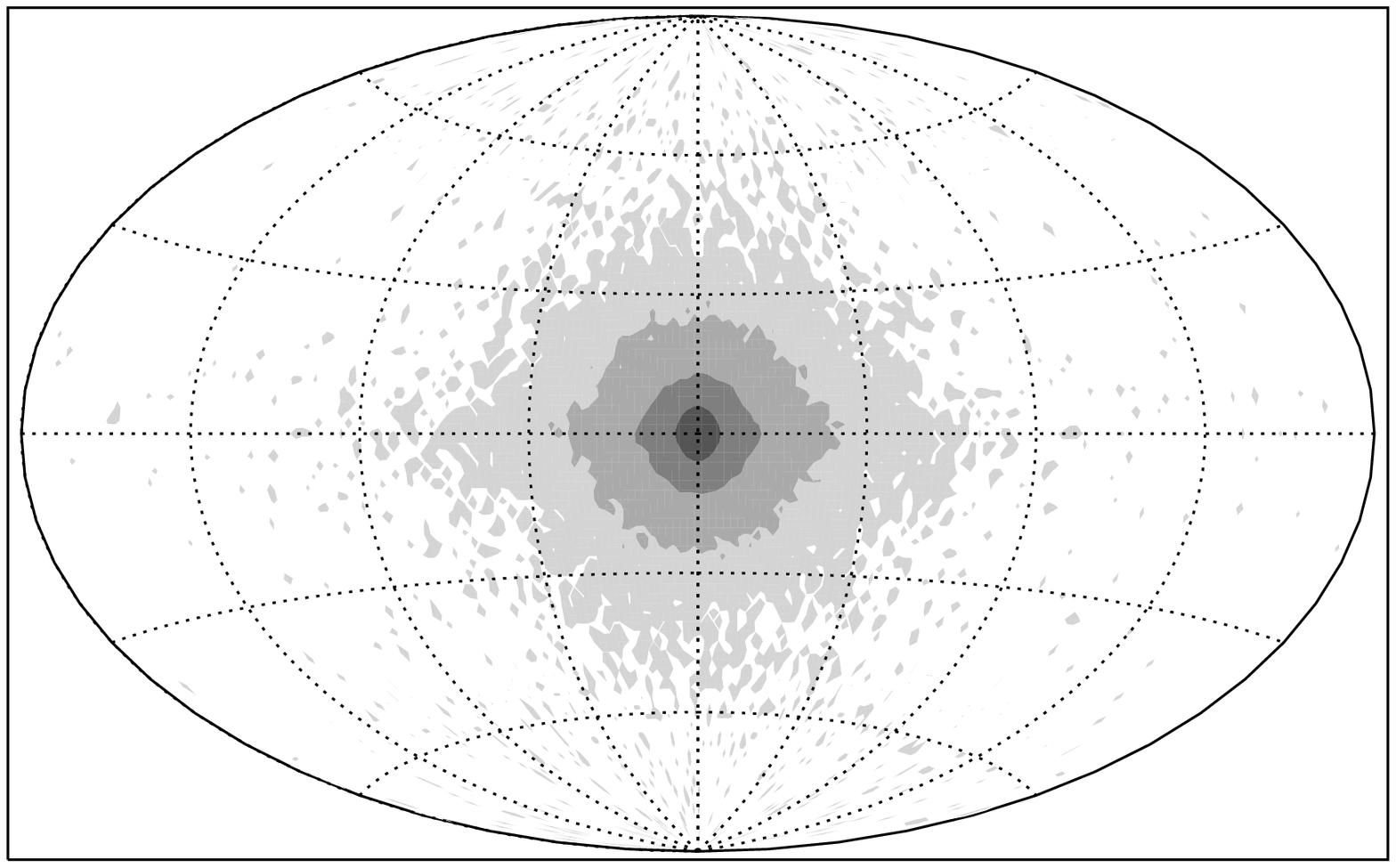}
  \includegraphics[width=0.45\textwidth]{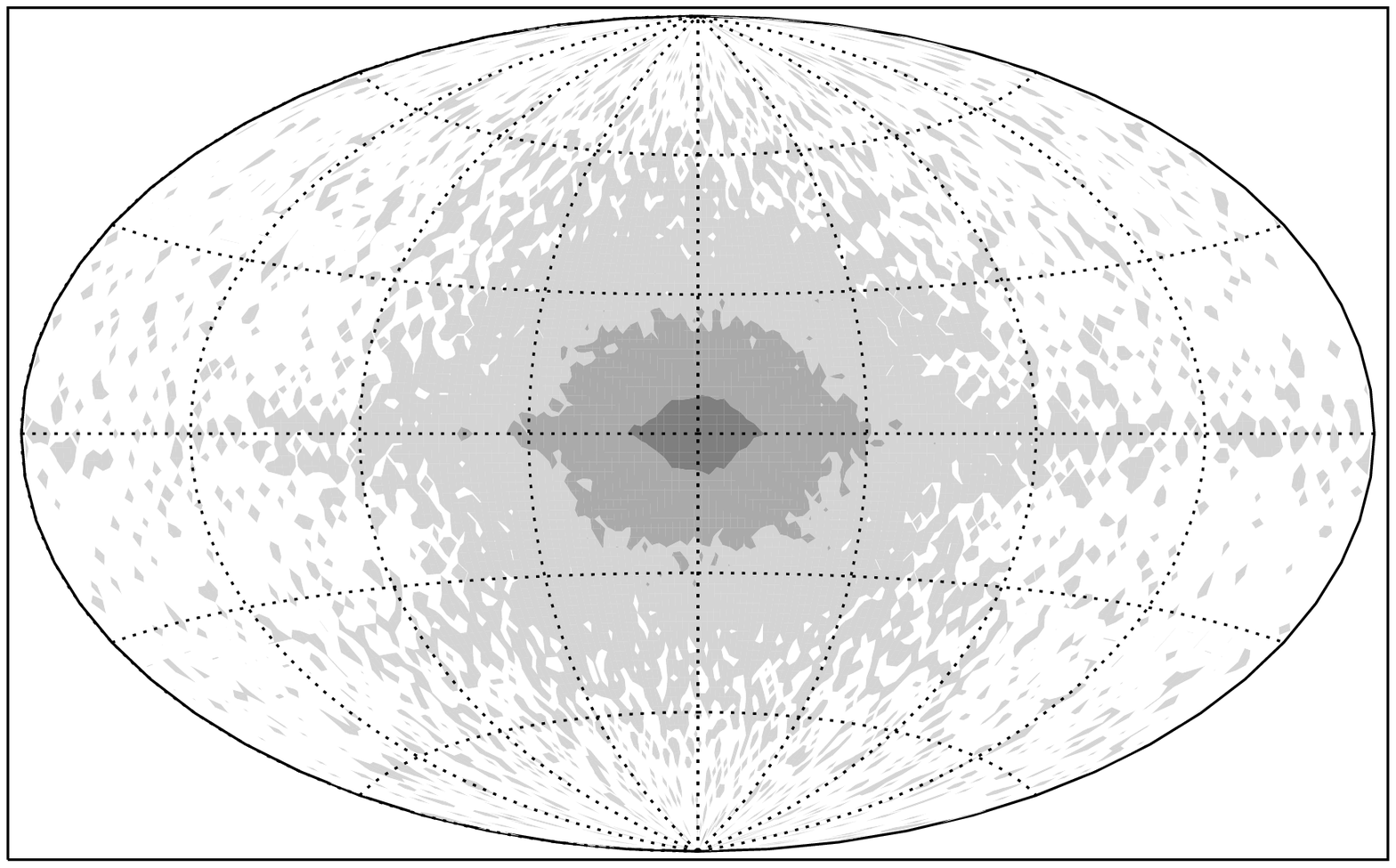}
	\includegraphics[width=0.45\textwidth]{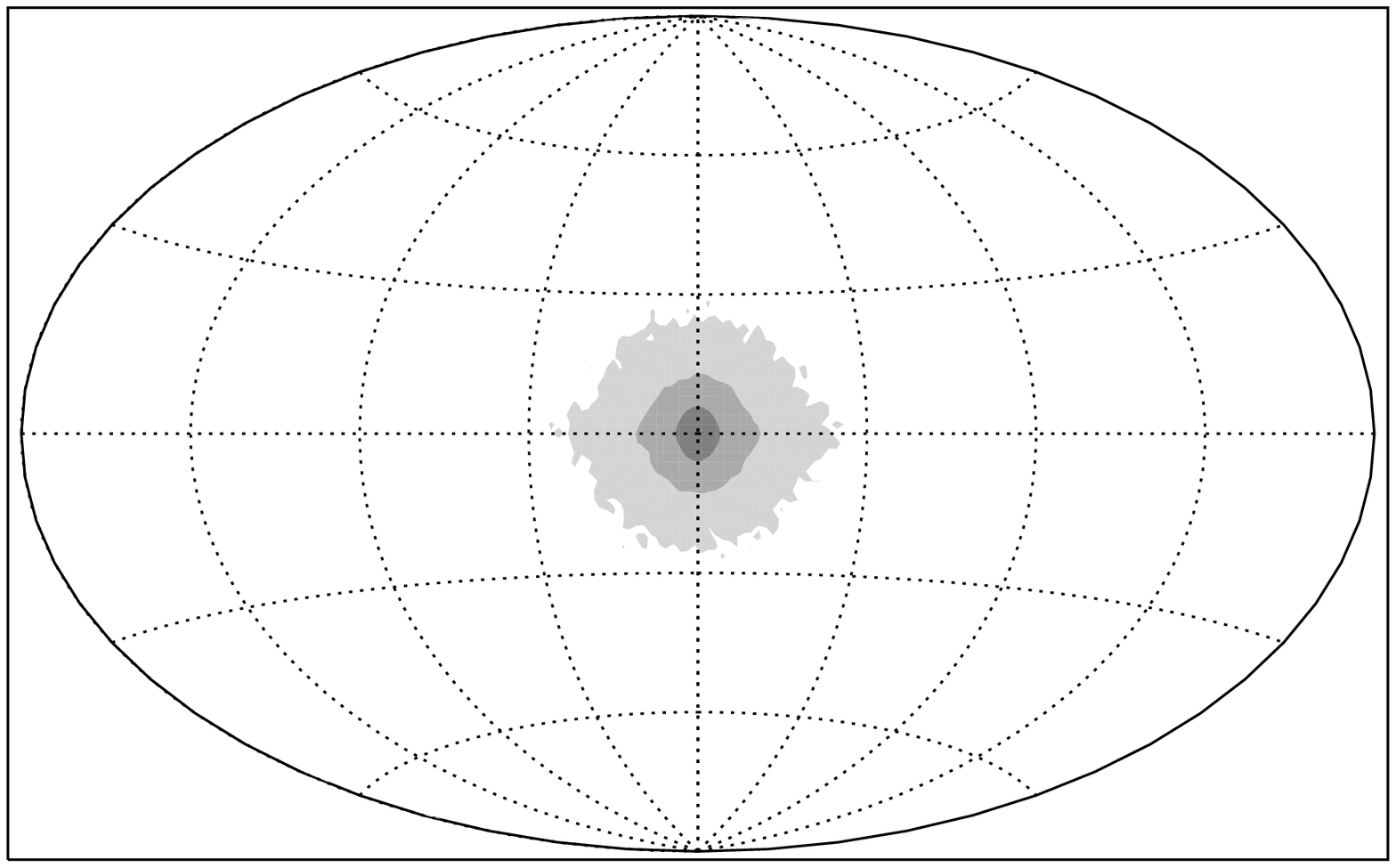}
  \includegraphics[width=0.45\textwidth]{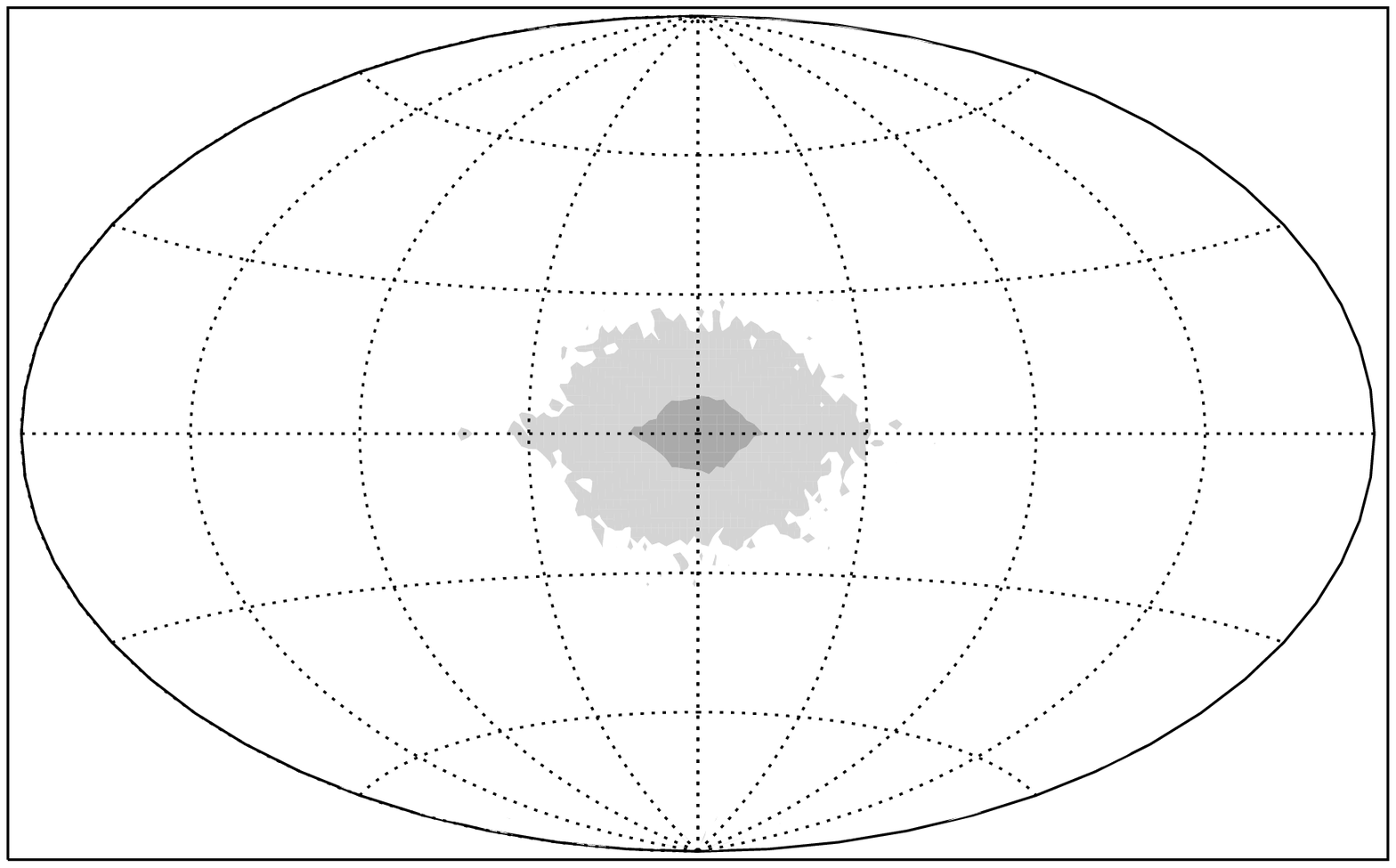}  
  \caption{Aitoff projection of the sky density of NSs (upper panels) and BHs (lower panels) within 12 kpc from the Sun, for bulge-born (left panels) and
  disk-born (right panels) objects. Contour profiles are drawn at $(10, 10^{2}, 10^{3}, 10^{4})\,\rm deg^{-2}$, respectively.
  Coordinates represent Galactic longitude (abscissae) and latitude (ordinates).}
  \label{pden}
\end{figure*}

Because we assumed that NSs and BHs have the same kinematic properties, the density ratio between BHs and NSs should be $\sim\,0.1$,
as expected from the ratio between the number fractions of the two populations (e.g. Table \ref{table_imf}).
Indeed we find that the local density of BHs is $\sim\,3.2\,\times\,10^{-6}\,\rm pc^{-3}$, while the density of BHs in the Galactic center is
$\sim\,1.1\,\times\,10^{-2}\,\rm pc^{-3}$. The fraction of bulge-born objects is the same of NSs.
The projected density of BHs in all directions is also $\sim\,10$ times lower than the projected density of NSs (Fig. \ref{pden}, lower panels).

\subsection{Optical depth}\label{results1}
We estimate the microlensing optical depth (Eq. \ref{eqn_tau2}) for different l.o.s. toward the Galactic bulge.
In particular we consider a $20°\,\times\,20°$ window, centered at ($l,b)=(0°,0°)$ and study the dependence of $\tau$ 
on the Galactic longitude and latitude.
We assume 100 percent detection efficiency in our calculations and do not take into account the effect of interstellar extinction nor do we apply any flux limit on source stars.
In this regard, we consider as sources only the stars in the main sequence, because we expect brown and white dwarfs to be too weak to be efficiently monitored by present surveys.

\begin{figure*}[t]
  \centering
  \includegraphics[width=0.49\textwidth]{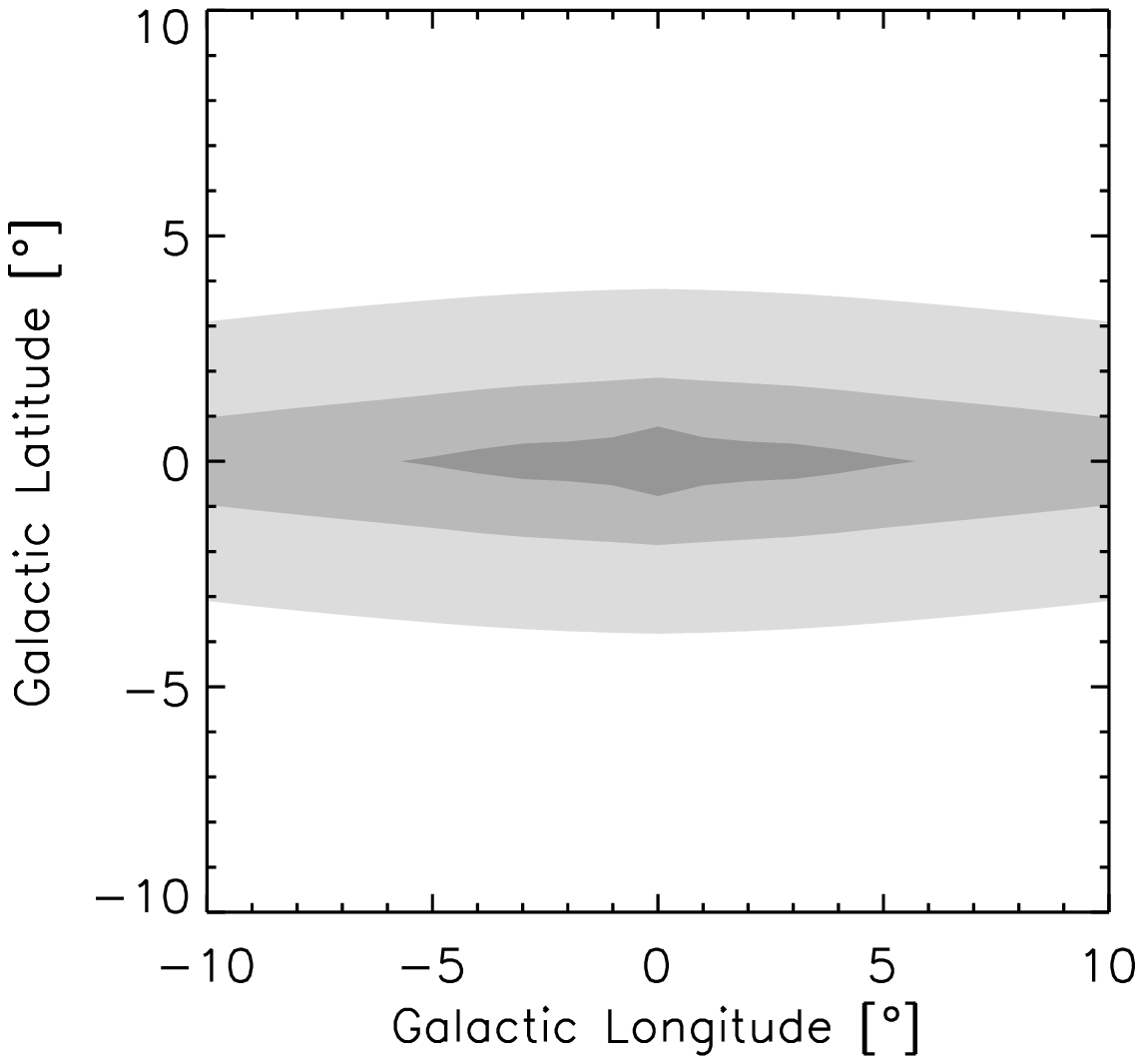}
  \includegraphics[width=0.49\textwidth]{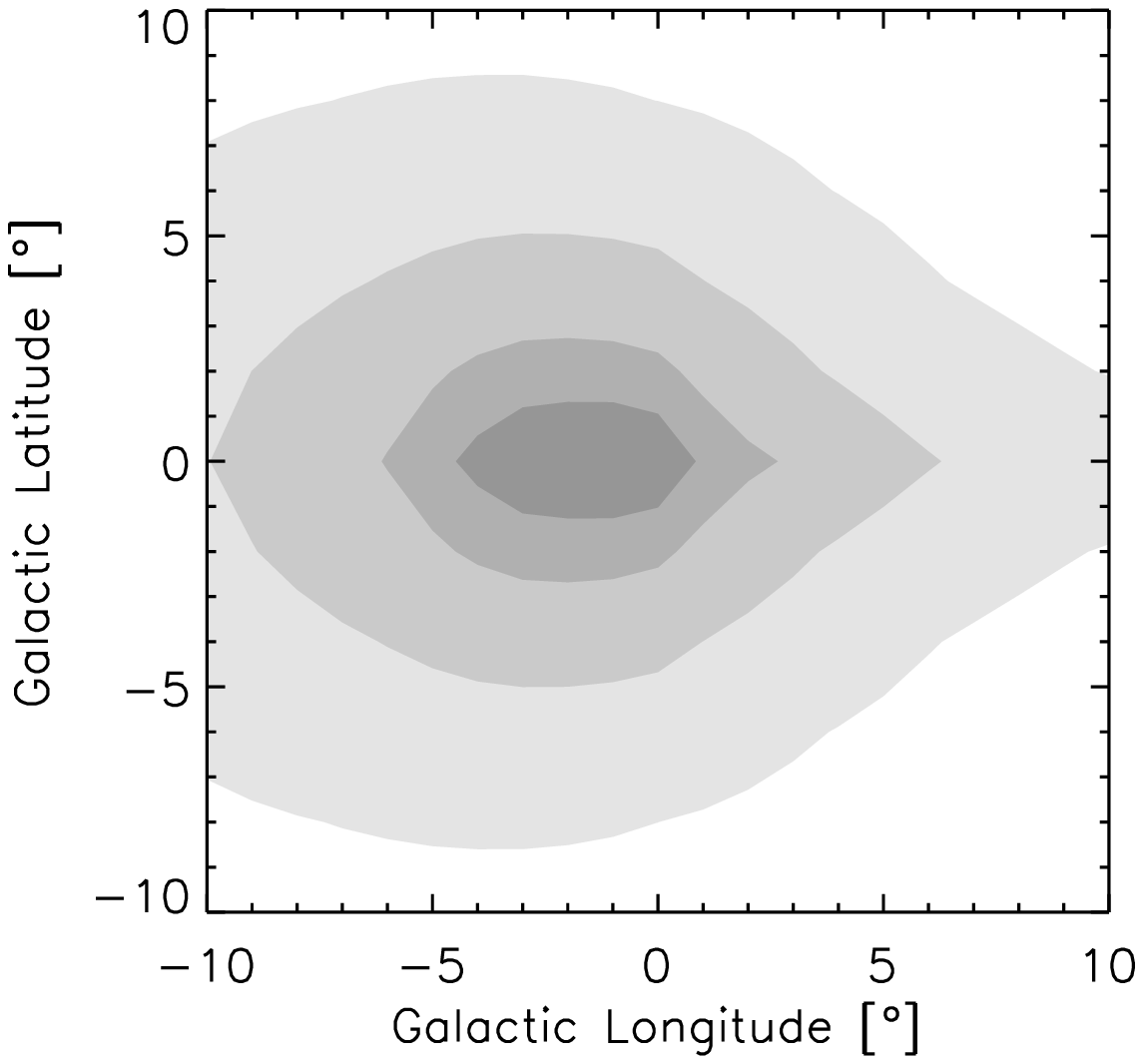}
  \caption{Optical depth profiles of normal stars (left) and ICoRs (right) as a function of the Galactic coordinates.
  Contour profiles are drawn at $(0.5, 1.0, 2.0, 3.0)\times10^{-6}$ and $(0.5, 1.0, 3.0, 6.0, 8.0)\times10^{-8}$.
  Darker gray levels indicate larger optical depth.}
\label{tau_maps}
\end{figure*}

In Fig. \ref{tau_maps} we show the contour plots of the optical depth of normal stars and ICoRs.
For normal stars the optical depth mostly depends on the Galactic latitude.
On the other hand, the dependence of $\tau_{star}$ from the longitude is weaker (compare this figure with Fig. 1 of \citealt{CN08}).
The optical depth of ICoRs instead has a strong dependence on both the longitude and latitude. 
Also, the asymmetry with respect to the Galactic center is more noticeable (Fig. \ref{tau_maps}, right panel).
The maximum of the optical depth is shifted toward negative longitudes, $l\,\sim\,-2°$.
This behavior is due to the l.o.s. intercepting the far end of the Galactic bar when looking toward negative longitudes.
In this condition the geometry of the source-lens system favors a larger Einstein radius, which implies a higher optical depth.

\begin{table}[ht]
	\centering
		\begin{tabular}{@{}c c c c c c }
\hline
l.o.s. & $\tau_{star}$ & $\tau^{(b)}_{NS}$ & $\tau^{(d)}_{NS}$ & $\tau^{(b)}_{BH}$ & $\tau^{(d)}_{BH}$ \\
$(l,b)$ & [$\times\,10^{-6}$] & [$\times\,10^{-8}$] & [$\times\,10^{-8}$] & [$\times\,10^{-8}$] & [$\times\,10^{-8}$] \\
\hline
$(1°, -3.9°)$ & 0.94 & 1.51 & 0.31 & 1.05 & 0.21 \\
$(1°.5, -2.68°)$ & 1.43 & 1.97 & 0.38 & 1.36 & 0.27 \\
$(1°.5, -2.75)°$ & 1.40 & 1.93 & 0.37 & 1.33 & 0.26 \\
\hline		
		\end{tabular}
\caption{Optical depth of normal stars, NSs and BHs toward different l.o.s. The superscripts \textit{(b)} and \textit{(d)} refer to the subpopulations of 
bulge and disk objects, respectively.}
\label{table_tau}
\end{table}

In Table \ref{table_tau} we report the optical depth of normal stars, NSs and BHs for several specific l.o.s in the selected window.
We find that the optical depth of normal stars is, for example, $\tau_{star}\sim0.94\times10^{-6}$
toward the Baade's Window, $(l,b)=(1°,-3°.9)$ while we obtain $1.43\times10^{-6}$ and $1.40\times10^{-6}$ toward $(l,b)=(1°.50,-2°.68)$ and $(l,b)=(1°.50,-2°.75)$. 
These values agree reasonably well with those found in the literature, see Sect. \ref{intro}.

Without dynamical evolution, the contribution of ICoRs to the optical depth would be proportional to the mass fraction of these objects.
Also, it would not depend on the line of sight.
From our assumptions on the IMF, the optical depth of NSs and BHs would be $\sim\,2.3$ and $\sim\,1.6$ percent of the total.

When the kinematics is taken into account, the optical depth differs from the value expected without dynamical evolution.
We find that the optical depth is slightly lower than the nominal value obtained when the kinematics is not accounted for. 
Also, we find that the contribution of ICoRs to the optical depth is mostly due to objects born in the bulge, $\tau^{(b)}/(\tau^{(d)} + \tau^{(b)})\,\sim\,0.84$, where $\tau^{(b)}$ and $\tau^{(d)}$ are the optical depths of bulge-born and disk-born ICoRs, respectively.
The contribution of BHs is a factor $\sim1.5$ lower of that of NSs.

\subsection{Event rates}\label{results2}

\begin{figure*}[htbp]
  \centering
    \includegraphics[width=0.49\textwidth]{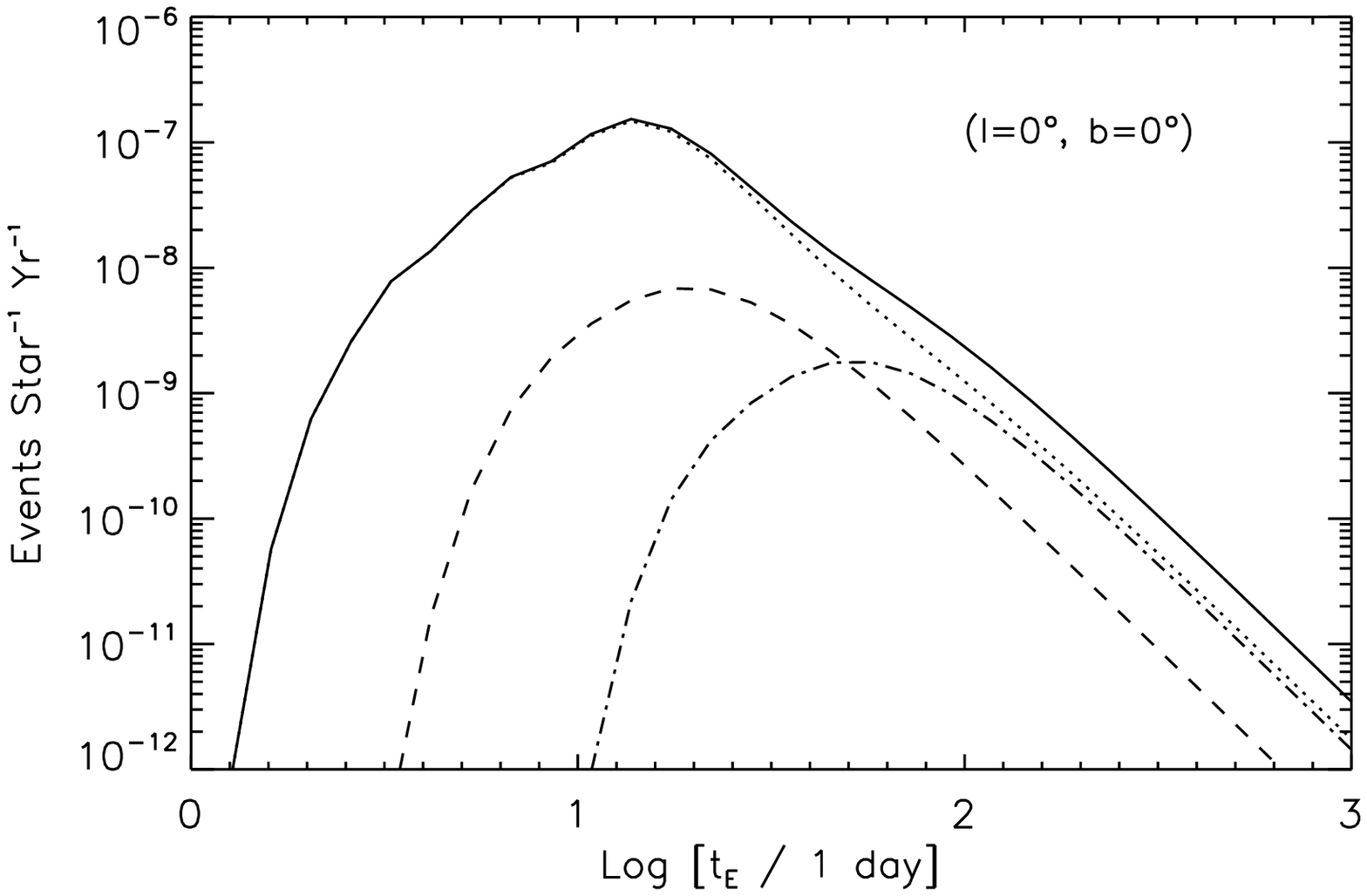}
			\includegraphics[width=0.49\textwidth]{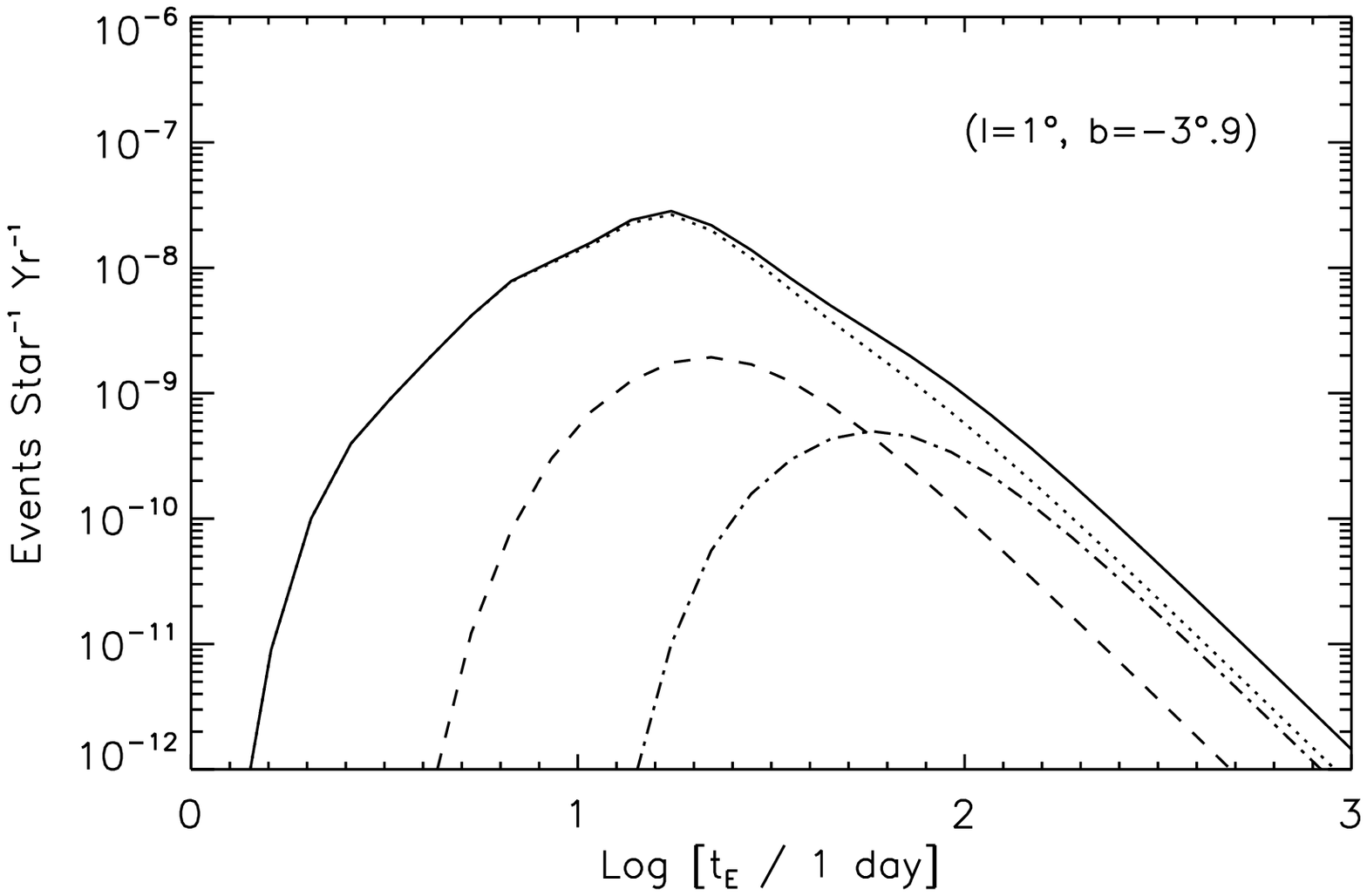}			
  \caption{Distribution of time scales for two different l.o.s., $(l,b)=(0°,0°)$ (left) and $(l,b)=(1°,-3°.9)$ (right).
 Each panel shows the total contribution (solid) and those of normal stars (dotted), NSs (dashed) and BHs (dot-dashed), respectively.}
\label{gamma_rates}
\end{figure*}

\begin{figure*}[htbp]
  \centering
 	\includegraphics[width=0.49\textwidth]{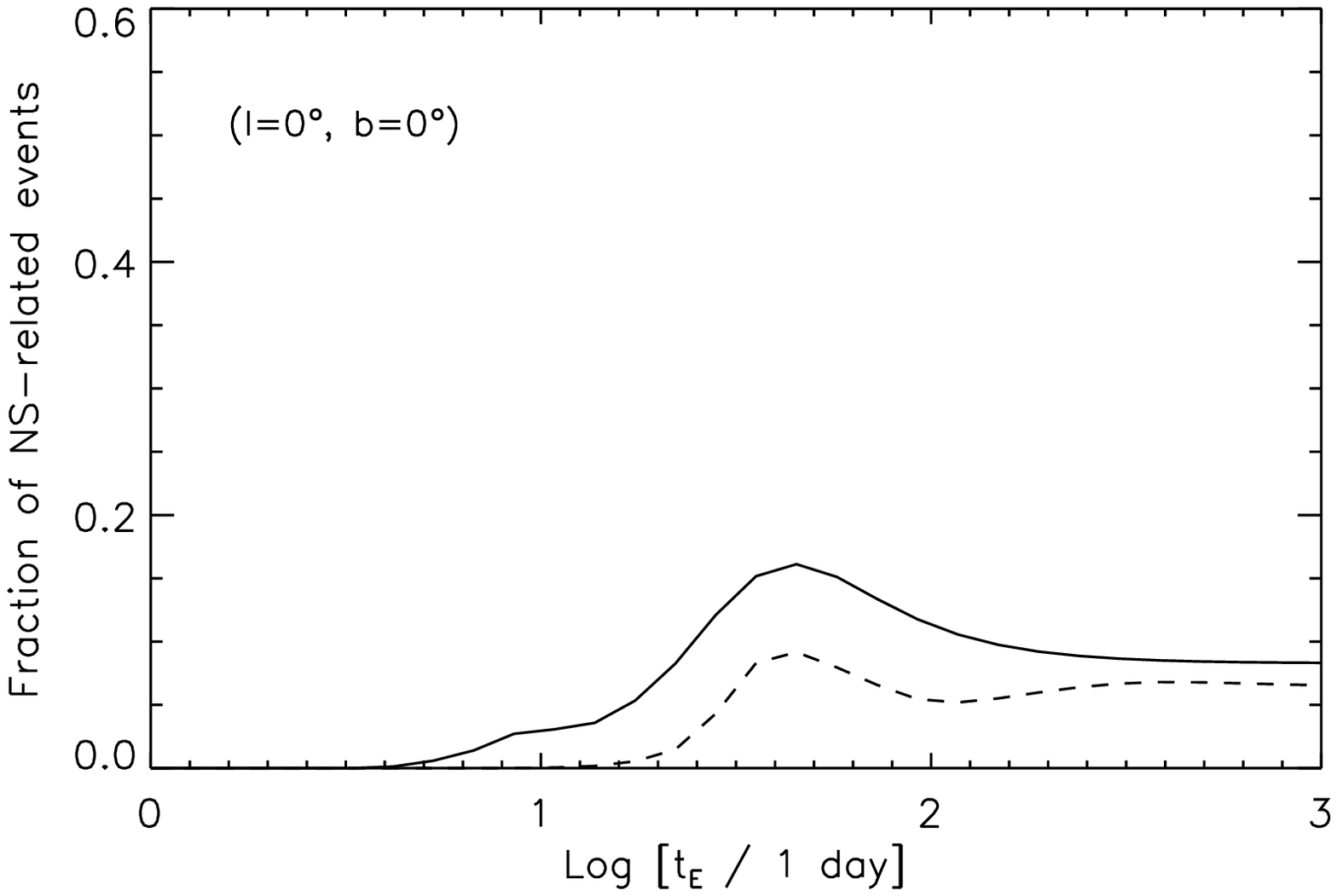}			
  		\includegraphics[width=0.49\textwidth]{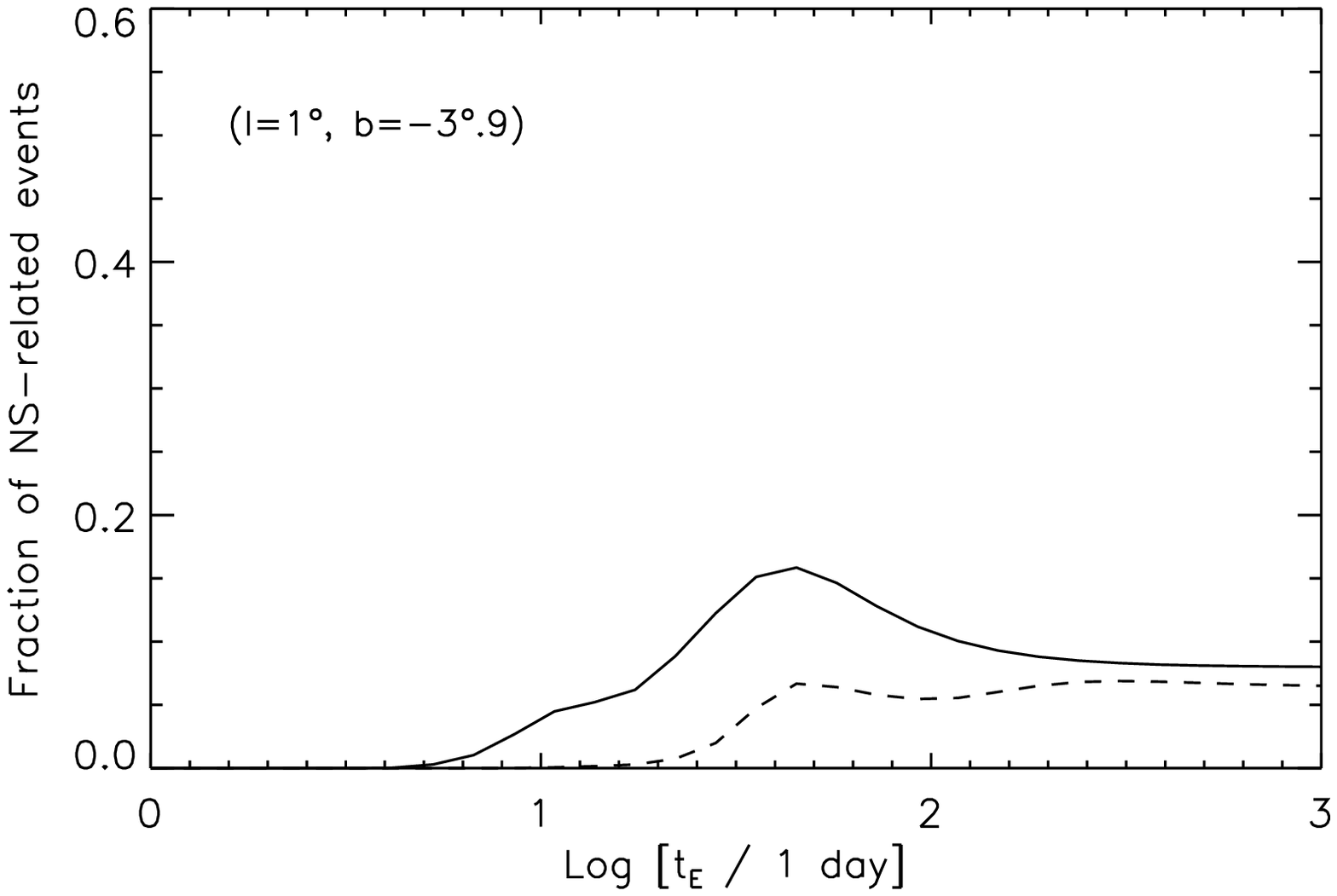}			
		\includegraphics[width=0.49\textwidth]{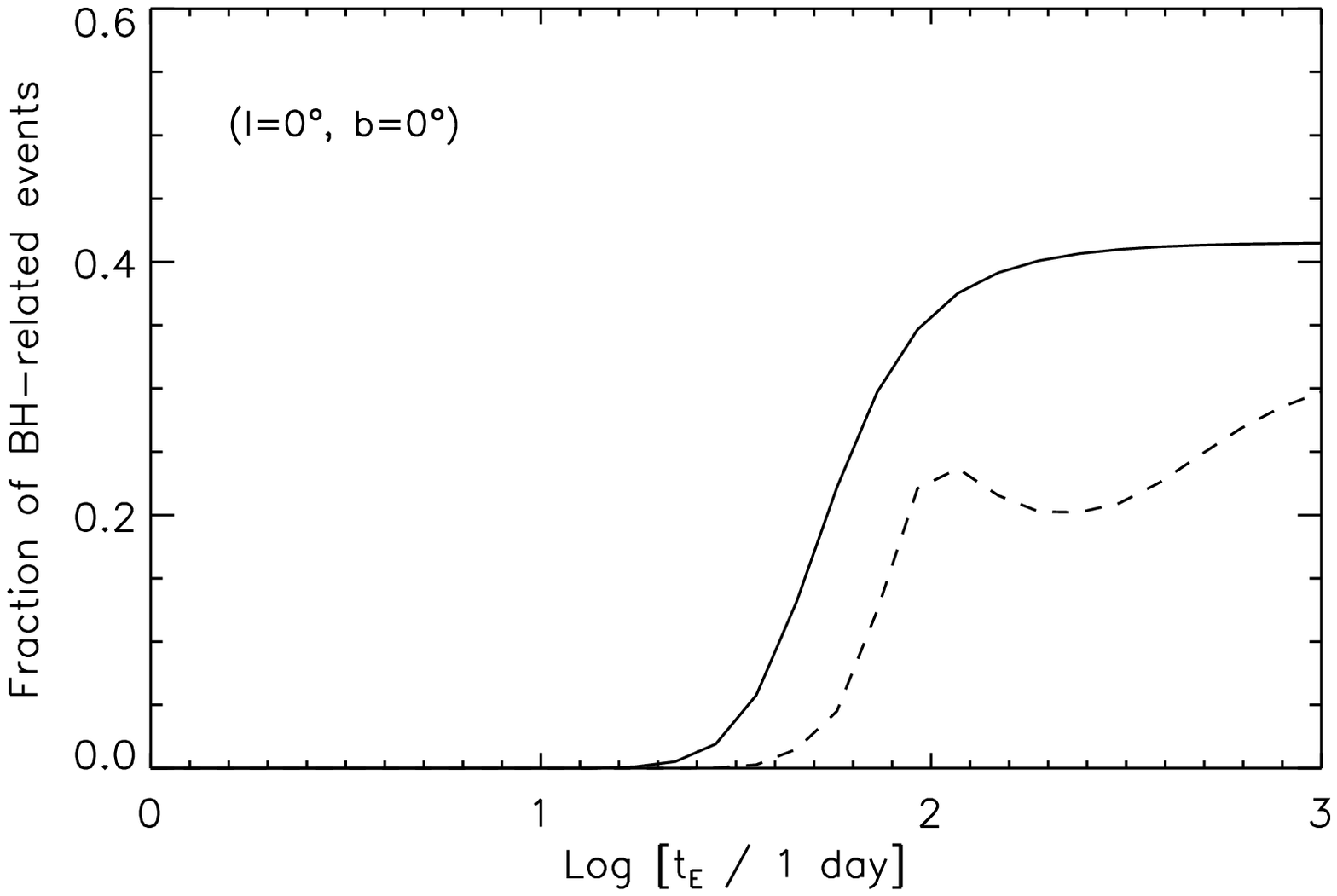}
     	\includegraphics[width=0.49\textwidth]{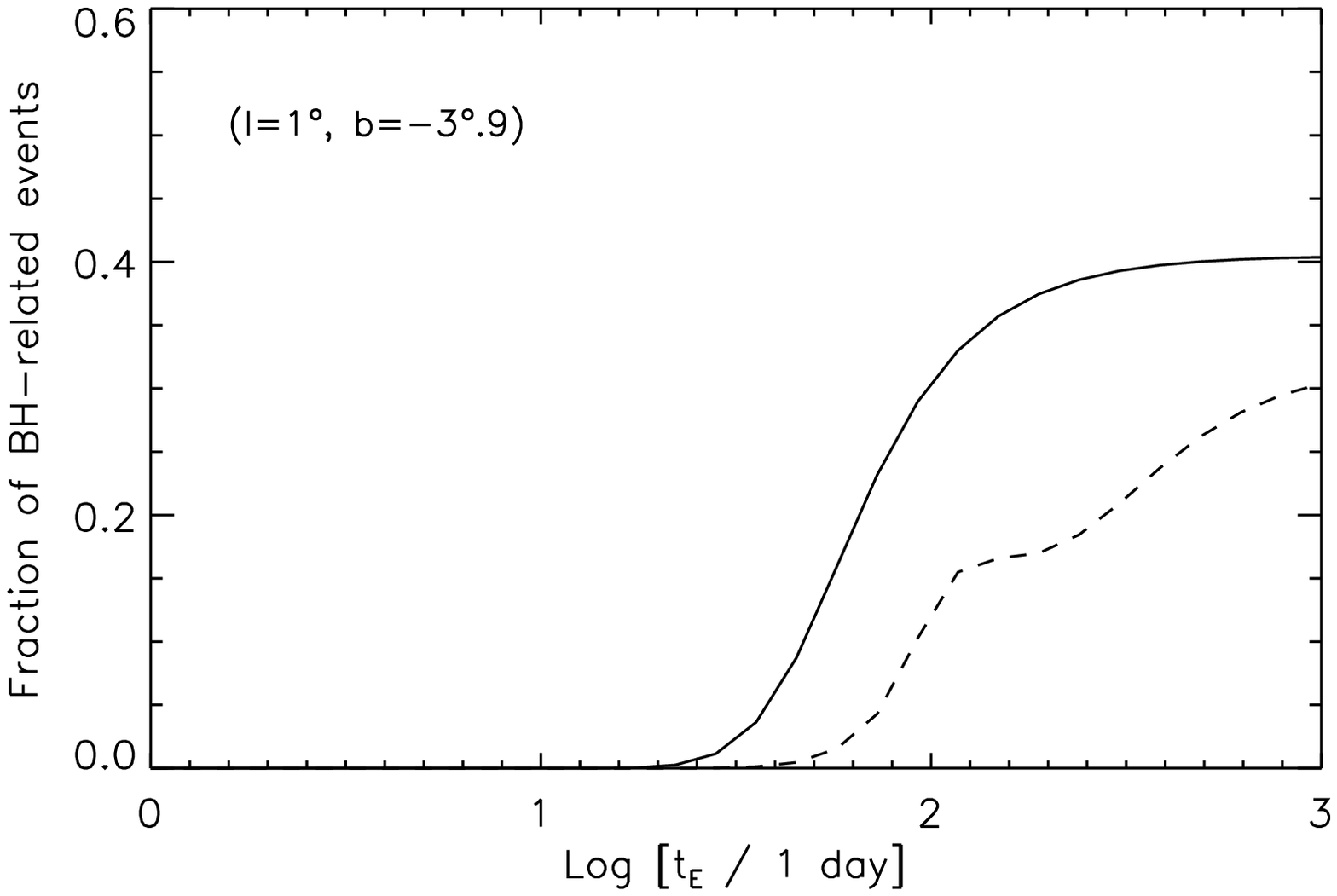}    
  \caption{Fractional contribution of ICoRs to the microlensing rate for NSs (left panels) and BHs (right panels).
The dashed line represents the expected contribution when kinematic effects are not taken into account.}
\label{gamma_fractions}
\end{figure*}

We calculated the expected distribution of event time scales for various l.o.s. toward the bulge.
Results for the Galactic center and the Baade's Window are reported in Table \ref{table_gamma} as examples.
In general, microlensing events are dominated by self-lensing of bulge low mass stars, these events having a typical duration of $\sim\,15\,-\,20$ days
(see e.g. Fig. \ref{gamma_rates}).
As expected the rate of events drops away from the plane (see Table \ref{table_gamma}).
The relative contribution of compact remnants is increased by a factor $\sim\,5$ with respect to the case with no kinematic effects.
Indeed, we find that the overall contribution of NSs rises from $\sim\,1$ to $\sim\,5$ percent, while for BHs it rises from $\sim\,0.2$ to $\sim\,1$ percent (Fig. \ref{gamma_fractions}).

The average duration of the events associated to remnants is lower by a factor $\sim\,1.5$, owing to the high velocities of these objects. 
For NSs $\left\langle t_E \right\rangle\,\sim\,25$ days instead of $\sim\,36$ days, while for black holes we find $\left\langle t_E \right\rangle\,\sim\,67$ instead of $\sim\,95$ days.
Intriguingly, the relative contribution of BHs increases with the time scale.
For events with duration longer than 100 days, this contribution is $\sim\,40$ percent (see Fig. \ref{gamma_fractions}), while NSs account for $\sim\,10$ percent of the events.

\begin{table*}[htpb]
	\centering
		\begin{tabular}{c c c c c c c}
\hline
l.o.s. & $\Gamma_{star}$ & $\left\langle t_E \right\rangle_{star}$ & $\Gamma_{NS}$ & $\left\langle  t_E \right\rangle_{NS}$ & $\Gamma_{BH}$ & $\left\langle t_E \right\rangle_{BH}$ \\
$(l,b)$ & [$10^{-5}\,\rm{star^{-1}\,yr^{-1}}$] & [days] & [$10^{-6}\,\rm{star^{-1}\,yr^{-1}}$] & [days] & [$10^{-6}\,\rm{star^{-1}\,yr^{-1}}$] & [days] \\ 
\hline
$(0°, 0°)$ & 2.67 & 16 & 1.47 & 25 & 0.38 & 67 \\
$(1°, -3°.9)$ & 0.52 & 20 & 0.40 & 28 & 0.10 & 77 \\
\hline		
		\end{tabular}
\caption{Rates and average time scales toward the Galactic center and the Baade's window.}
\label{table_gamma}
\end{table*}

\begin{table*}[htpb]
	\centering
		\begin{tabular}{c c c c c c c}		
\hline
l.o.s. & $f_{NS}$ & $f_{BH}$ & $f_{NS}$ & $f_{BH}$ & $f_{NS}$ & $f_{BH}$ \\
$(l,b)$ & All & All & ($1 < \log t_E < 2$) & ($1 < \log t_E < 2$) & ($\log t_E > 2$) & ($\log t_E > 2$) \\
\hline
$(0°, 0°)$ & 0.05 & 0.01 & 0.06 & 0.01 & 0.10 & 0.38 \\
$(1°, -3°.9)$ & 0.07 & 0.02 & 0.08 & 0.02 & 0.09 & 0.35 \\
\hline		
		\end{tabular}
\caption{Fractional contribution of NSs and BHs to the rate of events. The l.o.s selected are the same as in Table \label{table_gamma_fract}.}
\label{table_gamma_fract}
\end{table*}

\section{Discussion}\label{discussion}

Stimulated by recent results on the distribution of ONS in the Milky Way, we estimated the
contribution of isolated compact remnants (NSs and BHs) to the microlensing optical depth and event rate toward the Galactic bulge. 
We first ran new simulations of remnant orbits in a refined model of the 
Galactic gravitational potential, which accounts for the observed distribution of stars obtained from IR surveys. 
In our model, the Galactic bulge accounts for $\sim\,45$ percent of the compact remnants born in the MW.
Our results show that the initial conditions play a fundamental role in the final fate of an ICoR.
At variance with disk-born objects, almost all of those born in the bulge are likely retained by the Galaxy, and thus they give the major contribution
to the microlensing rate toward the bulge itself.

We found that the microlensing optical depth of ICoRs is lower than that obtained without dynamical evolution 
(i.e. no kick at birth).
On the other hand, the net effect of the high velocities on microlensing is an increase of the event rate and a decrease in the typical time scale of the events. 
Indeed, we found that the contribution to the event rate is higher than without kinematics effects.
The average time scale of events associated to ICoRs is also lower than that expected from a population of remnants without kicks.
Interestingly, for durations $\gtrsim\,100$ days, $\sim30-40$ percent of the events observed may indeed be related to BHs.

Up to date, thousands of events have been observed toward the Galactic bulge by the various surveys \citep{Mo10}. 
Thus our results suggest that at least several hundreds of events related to ICoRs could be present in the catalogs. 
These are likely to be hidden among long-duration events.
An excess of long-duration events has indeed been reported by \cite{Po05}.
These event are likely to be generated by massive objects, like NSs and BHs.

The basic problem is therefore to discuss a procedure to distinguish isolated compact remnants from normal stars that are responsible for the microlensing events.
While this discussion is out of the scope of the present paper, it is worth to add some considerations.
The only source of steady luminosity for ICoRs suggested thus far is accretion from the ISM (e.g \citealt{ORS70}, \citealt{TC91}, \citealt{BM93}, \citealt{AK02}).
However, in the case of NSs, the absence of sources in the accreting phase as clearly demonstrated by the ROSAT surveys \citep{NT99}
indicates that accretion is inhibited most probably by the rotating magnetic structure (e.g. \citealt{To03}, \citealt{Pe03}).
Accretion onto isolated BHs is favored with respect to the NSs, because they are more massive and do not have a magnetic field that can hamper the accretion flow.
However, they are far less numerous than NSs.

These considerations should be revisited, because we now consider the remnants in the bulge.
The magnetic field of NSs and the properties of the ISM are different from what was discussed so far (e.g. \citealt{Za96}).
One should first calculate a realistic X-ray luminosity (see e.g. on this line \citealt{BP10} and references therein), corrected for absorption, 
and then compare with present and future X-ray missions.
The advantage with respect to a blind search of ICoRs is that microlensing events give a precise location of the object.

We note that, if our calculations are correct, microlensing could be the only way to probe the velocity distribution of isolated BHs, which has not yet been  constrained.

\begin{acknowledgements}
We thank the referee for suggesting a procedure to check our calculations, which allowed us to find an error in a previous version of our code.
We also thank P. Jetzer, R. Turolla, and R. Salvaterra for helpful discussion.
\end{acknowledgements}


\end{document}